\documentclass[aps,prd,twocolumn,groupedaddress,nobibnotes,superscriptaddress,nofootinbib,10pt]{revtex4-1}

\usepackage{times}
\usepackage{amsmath}
\usepackage{graphicx}
\DeclareGraphicsExtensions{.pdf,.eps,.png,.jpg}
\usepackage{acronym}
\usepackage{color}
\usepackage[caption=false]{subfig}
\usepackage{tabularx}
\usepackage{longtable}
\usepackage{multirow}
\usepackage{dcolumn}
\usepackage{soul}
\usepackage{makecell}
\usepackage{float}
\usepackage{hyperref}
\usepackage{cleveref}
\usepackage{gensymb}
\usepackage[usenames,dvipsnames,svgnames]{xcolor}

\newcolumntype{L}[1]{>{\raggedright\let\newline\\\arraybackslash\hspace{0pt}}m{#1}}
\newcolumntype{C}[1]{>{\centering\let\newline\\\arraybackslash\hspace{0pt}}m{#1}}
\newcolumntype{R}[1]{>{\raggedleft\let\newline\\\arraybackslash\hspace{0pt}}m{#1}}

\begin{document}

\title{Deep Learning Model on Gravitational Waveforms in Merging and Ringdown Phases of Binary Black Hole Coalescences}

\author{Joongoo Lee}
\email{ljg4471@gmail.com}
\affiliation{Korea Astronomy and Space Science Institute, 776 Daedeokdae-ro, Yuseong-gu, Daejeon 34055, Republic of Korea}
\affiliation{Department of Physics and Astronomy, Seoul National University, Seoul 08826, Republic of Korea}
\author{Sang Hoon Oh}
\email{shoh@nims.re.kr}
\affiliation{Division of Basic Researches for Industrial Mathematics, National Institute for Mathematical Sciences, Daejeon 34047, Republic of Korea}
\author{Kyungmin Kim}
\affiliation{Korea Astronomy and Space Science Institute, 776 Daedeokdae-ro, Yuseong-gu, Daejeon 34055, Republic of Korea}
\author{Gihyuk Cho}
\affiliation{Deutsches Elektronen-Synchrotron DESY, Notkestrasse 85, 22607 Hamburg, Germany}
\author{John J. Oh}
\affiliation{Division of Basic Researches for Industrial Mathematics, National Institute for Mathematical Sciences, Daejeon 34047, Republic of Korea}
\author{Edwin J. Son}
\affiliation{Division of Basic Researches for Industrial Mathematics, National Institute for Mathematical Sciences, Daejeon 34047, Republic of Korea}
\author{Hyung Mok Lee}
\affiliation{Korea Astronomy and Space Science Institute, 776 Daedeokdae-ro, Yuseong-gu, Daejeon 34055, Republic of Korea}
\affiliation{Department of Physics and Astronomy, Seoul National University, Seoul 08826, Republic of Korea}

\begin{abstract}
The waveform templates of the matched filtering-based gravitational-wave search ought to cover wide range of parameters for the prosperous detection. Numerical relativity (NR) has been widely accepted as the most accurate method for modeling the waveforms. Still, it is well-known that NR typically requires a tremendous amount of computational costs. In this paper, we demonstrate a proof-of-concept of a novel deterministic deep learning (DL) architecture that can generate gravitational waveforms from the merger and ringdown phases of the non-spinning binary black hole coalescence. Our model takes ${\cal O}$(1) seconds for generating approximately $1500$ waveforms with a 99.9\% match on average to one of the state-of-the-art waveform approximants, the effective-one-body. We also perform matched filtering with the DL-waveforms and find that the waveforms can recover the event time of the injected gravitational-wave signals.
\end{abstract}

\maketitle

\section{Introduction}
Since the first detection of gravitational waves (GW)\cite{Abbott:2016blz}, numerous GW events have been captured by ground-based GW detectors, the Advanced Laser Interferometer Gravitational-wave Observatory (aLIGO)~\cite{TheLIGOScientific:2014jea} and Virgo~\cite{TheVirgo:2014hva}. The sources of all events turned out to be compact binary coalescences (CBCs), the collision of two dense objects such as black holes (BH) or neutron stars (NS) --- mostly from binary black holes (BBH), 47 out of 50, and partially from binaries containing at least one neutron star~\cite{2020arXiv201014527A}.

For the type of GW progenitors, template-based GW search is one of the most efficient approaches because the gravitational waveforms from binary mergers can be modeled precisely by multiple methods, e.g., post-Newtonian (PN) for the inspiral phase, numerical relativity for the merger phase, and perturbation theory for the ringdown phase.  The template-based search utilizes the matched filtering method~\cite{1057571}, which essentially computes the cross-correlation between template waveforms and real GW signal buried in noisy data.

The successful implementation of the matched-filtering-based search relies on the pre-computed waveform templates. Numerical relativity (NR) has been considered as the most accurate method for computing gravitational waveforms. However, obtaining a large number of templates that cover parameter space densely enough for the precise matched filtering search and parameter estimation with NR is not feasible because of too heavy computational requirements. For example, NR simulation of the first GW event GW150914~\cite{Abbott:2016blz} takes 1-2 weeks using tens to hundreds of CPU cores~\cite{Lovelace:2016uwp}. In contrast, it takes less than $\mathcal{O}(1)$ seconds to generate inspiral waveforms using post-Newtonian approximations.

Several waveform models approximating NR waveforms have been proposed to reduce the computational cost with reasonable accuracy NR~\cite{Blanchet:1995ez,Droz:1999qx,Buonanno:2000ef,Blanchet:2004ek,Purrer:2014fza,Taracchini:2013rva,Purrer:2015tud,bohe2016improved}. Nonetheless, the physical parameter spaces where each approximant exactly covers are different from each other~\cite{PhysRevD.93.104050,Taracchini:2013rva}.
Therefore, reserving plural waveform models, complementing each other for various configurations, and saving computing time are crucial for a more elaborate template-based search. It justifies the further study of new waveform approximants. 

We present a proof-of-concept demonstration of a deep learning (DL) model for generating gravitational waveforms from the CBC events covering the late phase of inspiral to final ringdown phases. For this purpose, we only consider non-spinning BBH systems for simplicity. Chua et al.~\cite{PhysRevLett.122.211101} utilize deep artificial neural networks to map the physical parameters to coefficients of reduced-order bases waveforms. Williams et al.~\cite{unknown} use Gaussian process regression to approximate the inspiral-merger-ringdown waveforms from the BBH. However, the capability of a fully DL-based deterministic approach has not been explored so far for the generation of the merger-ringdown waveform of CBC\footnote{It is known that deterministic models generally show higher accuracy and performance than stochastic methods as the training data is sufficient.}. Hence, we examine the viability of the deterministic DL model as a merger-ringdown gravitational waveform model throughout this paper.

While DL models show remarkable performances in a wide variety of fields such as natural language processing (NLP)~\cite{2020arXiv200514165B,2017arXiv170603762V}, autonomous driving~\cite{Dosovitskiy17}, and image classification~\cite{10.1145/3065386}, most of the models are only capable of handling fixed-size data once they are trained. However, the model we shall adopt for this study should be able to cope with differently-sized data because the length of the waveforms observable by GW detectors depends on the two factors: (1) lower-frequency limit of the detector's sensitivity (around 10 Hz for ground-based detectors) and (2) the masses of the compact binary system~\cite{TheLIGOScientific:2014jea,2020arXiv201014527A}.

\begin{table*}[t!]
  \caption{\label{Tab:data_info}Parameters of the waveform in each dataset. Dataset-1 and-2 have different mass ranges, mass ratios, and numbers of samples, as shown in the table. All the other parameters of both datasets are set to be the same. Note that the waveforms in the datasets are generated in the time domain with \texttt{PyCBC} and \texttt{SEOBNRv4}.}
  \centering
  \begin{tabularx}{\linewidth}{X c @{\hskip 1in} c}
    \toprule
    Variable & Dataset-1 & Dataset-2 \\ 
    \colrule
    Mass [min, max] & [10$M_\odot$, 40$M_\odot$] & [40$M_\odot$, 100$M_\odot$] \\ 
    Mass ratio [min, max] & [1, 4] & [1, 2.5] \\
    {Number of waveforms}(training, validation, test) & (12469, 1533, 1512) & (12447, 1530, 1523) \\
    sampling rate & 4096Hz & 4096Hz  \\ 
    Distance & 100Mpc & 100Mpc  \\ 
    Spin & 0 & 0  \\ 
    Inclination angle & 30\degree & 30\degree  \\ 
    \botrule
  \end{tabularx}
\end{table*}

The recurrent neural network (RNN) encoder-decoder-based sequence-to-sequence (seq2seq) model~\cite{cho2014learning,Sutskever:2014:SSL:2969033.2969173} designed for NLP is one of the DL models that can handle variable input/output sizes. This model also has shown outstanding performances in many NLP studies~\cite{gehring2017convolutional,venugopalan2015sequence,luong2015multi,nallapati2016abstractive}. The property of gravitational waveforms is similar to that of language type data containing time-ordered words in sentences with different lengths. In that sense, we consider seq2seq as the experimental method to generate waveforms and slightly modify the structure of the model for our purpose.

This paper is organized as follows. Sec.~\ref{sec:data_prep} provides detailed explanations on the data preparation. In Sec~\ref{sec:methods}, the original seq2seq model, our modified version, and an evaluation metric for the model performance are elaborated. Sec.~\ref{sec:results} presents the results of the DL-waveform analysis with GW data and additional dataset-size-associated experiments. Finally, we discuss our results and future research directions in Sec.~\ref{sec:discussion}.

\section{Data}
\label{sec:data_prep}

\begin{figure}[!t]
  \centering
    \includegraphics[width=1\linewidth,height=0.2\textheight]{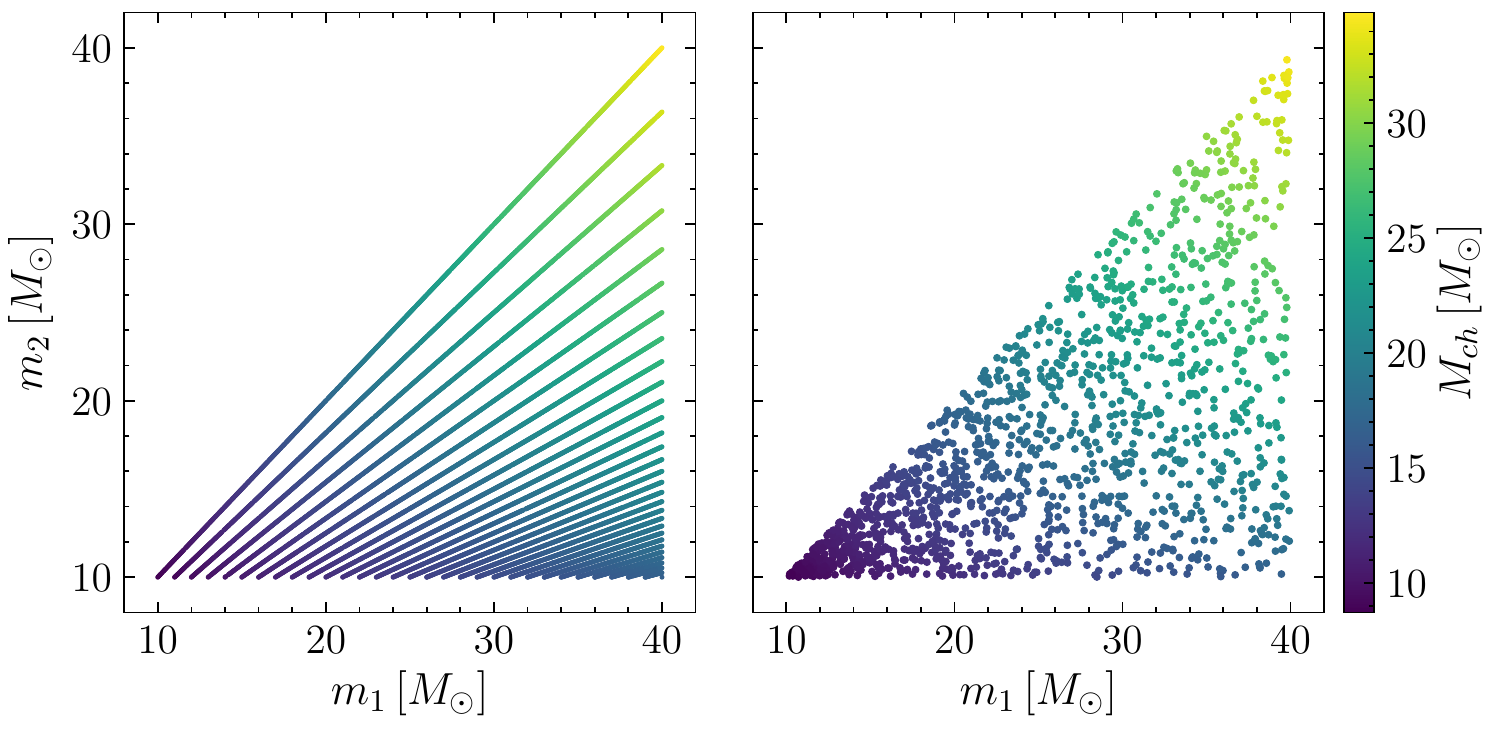}\\
    \includegraphics[width=1\linewidth,height=0.2\textheight]{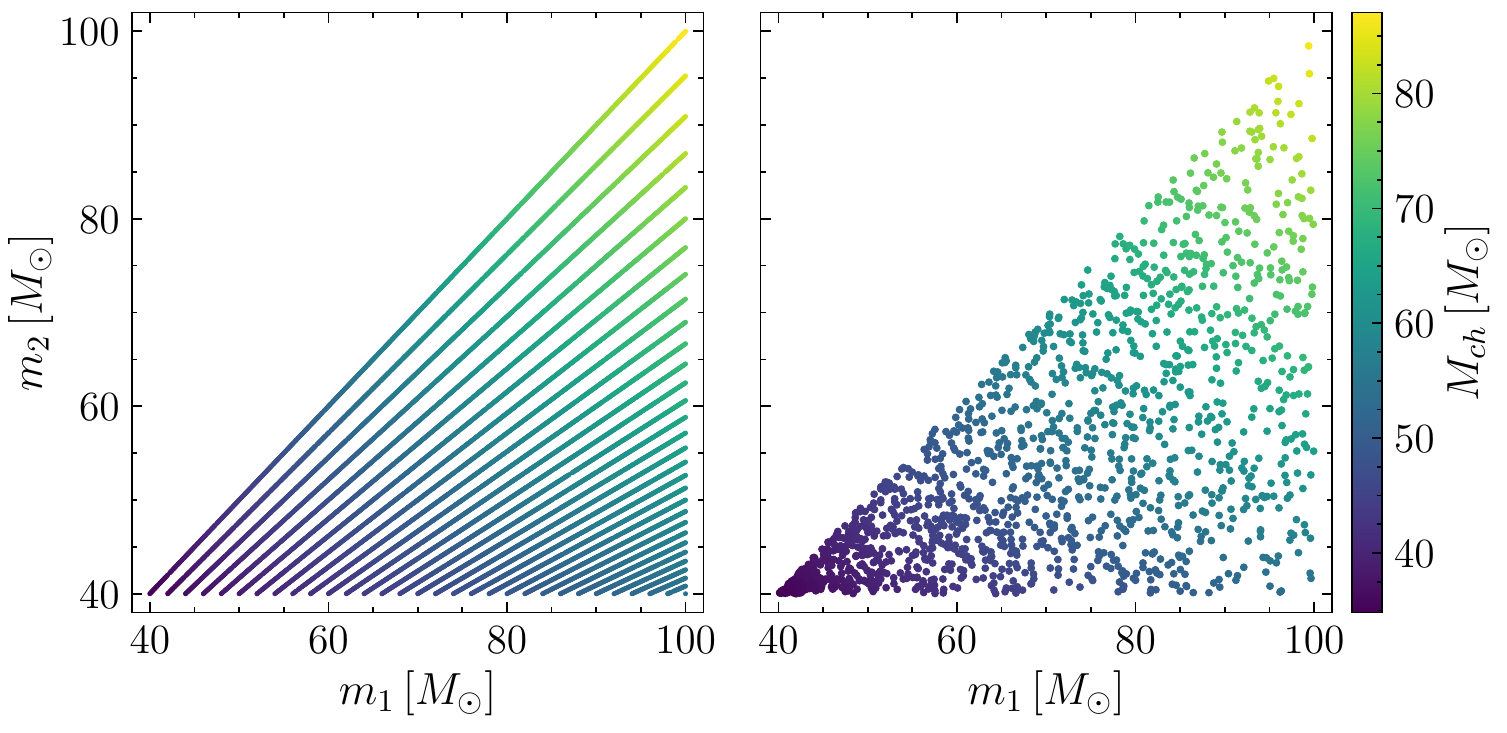}
    \vspace{-0.5cm}
  \caption{\label{Fig:scatter_contour}The component masses of training (left) and test (right) sub-datasets in dataset-1 (upper) and dataset-2 (bottom) with the color-coded chirp mass. While we use a set of fixed mass ratios, $m_1/m_2$, for the training sub-dataset, $m_1$ and $m_2$ are randomly chosen for the test sub-dataset with the restriction that $m_1 \ge m_2$. The mass ratios range from 1 to 4 for the dataset-1 and from 1 to 2.5 for dataset-2.}
\end{figure}

Since RNN is well-suited to time-series data, we compute non-spinning BBH waveforms in time-domain for training dataset using \texttt{PyCBC}~\cite{alex_nitz_2018_1490104}, a software package for GW data analysis. For this, we use a variant of effective-one-body (EOB) approximants, \texttt{SEOBNRv4}~\cite{PhysRevD.95.044028}, one of the most accurate versions of the approximants used in the GW searches.

For the training of the DL model, adopting waveforms obtained by NR is more beneficial than using approximants in the sense of accuracy. However, we find that the number of publicly available NR-waveforms of BBHs is only $\mathcal{O}(10^3)$~\cite{Boyle:2019kee,healy2019second,Jani_2016,Healy_2017}. In specific, the number of non-spinning BBH waveforms reduces to $\mathcal{O}(10^2)$~\cite{Boyle:2019kee}, so small that it might cause overfitting of the DL model~\cite{Grone:2017:HML:3153997}, which infects the general performance of the model. Thus we use EOB-waveforms to get a sufficient amount of training samples.

With the software and the approximant, we configure two datasets whose mass ranges of single black holes are [$10M_{\odot}$, $40M_{\odot}$] (dataset-1) and [$40M_{\odot}$, $100M_{\odot}$] (dataset-2) to divide search regions into low- and high-mass regions. Each dataset is consist of training, validation, and test sub-datasets with respective sample number ratio of 0.8, 0.1, and 0.1. The mass ratios of the sub-datasets are set differently\footnote{The mass ratio is defined as $m_{1}/m_{2}$, and $m_1 \geq m_2$ is assumed by convention.}. For the training and validation samples, we use fixed mass ratios with an interval of 0.1 (0.05) within the range of [1, 4] ([1, 2.5]) for dataset-1 (dataset-2). On the other hand, we randomly sample $m_1$ and $m_2$ in the corresponding parameter space for the test sub-dataset. In this manner, we can prove that the model trained with a limited mass ratio samples can be applied to the ones residing in any regions of the parameter space. Fig.~\ref{Fig:scatter_contour} shows the scatter plots of $m_1$ and $m_2$ of training sub-dataset in dataset-1 and -2 with color-coded chirp masses defined as $M_{ch}= (m_1 m_2)^{3/5}(m_1+m_2)^{-1/5}$. We use the sampling rate, distance, and inclination angle of 4096Hz, 100Mpc, and 30\degree, respectively. The parameters employed for data preparation are tabulated in Table~\ref{Tab:data_info}.

Following the data generation, the waveforms in dataset-1 and -2 are normalized with the maximum strain amplitude of each dataset. Since the diverse range of samples may cause biased results~\cite{589532}, data normalization for the differently ranged dataset is necessary. By normalizing the dataset, the sample values can be restricted in a comparable range and contribute equally to the DL model optimization at the beginning of the training.

\begin{figure}[t]
\centering
\includegraphics[width=.9\linewidth]{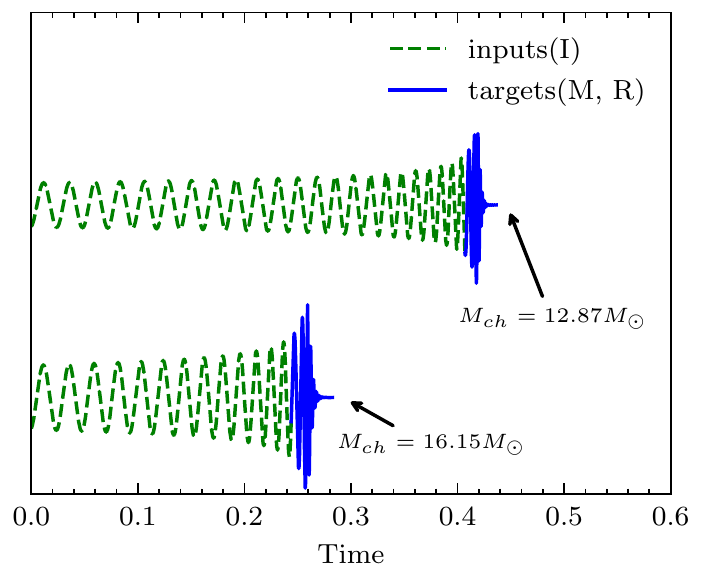}
\caption{Examples of input (green dashed; inspiral) and target (blue solid; merger-ringdown) waveforms drawn with different chirp masses of the compact binary system. They are computed by using \texttt{SEOBNRv4}. The upper and lower waveforms depict $M_{ch}=12.87M_{\odot}$ and $M_{ch}=16.15M_{\odot}$, respectively. Note that the length of the generated waveforms changes depending on the mass.}
\label{Fig:waveform}
\end{figure}

In turn, we divide each waveform into input and target waveforms: the input with the inspiral phase and target with merger and ringdown phases, respectively. For the division, we consider the point that the GW frequency reaches the innermost stable circular orbit~(ISCO) frequency~\cite{doi:10.1002/asna.19752960110} as the termination point of the inspiral phase~\cite{PhysRevD.83.024028}. The final data point of the input waveform and the initial data point of the target waveform are intentionally superposed to check whether the DL-waveform and given inspiral waveform are smoothly connected. Fig.~\ref{Fig:waveform} illustrates examples of input and target waveforms with different chirp masses. For the training of our DL model, we feed the input waveform to the DL model and let the model recover target waveform.

\begin{figure}[t]
 \includegraphics[width=1.\linewidth]{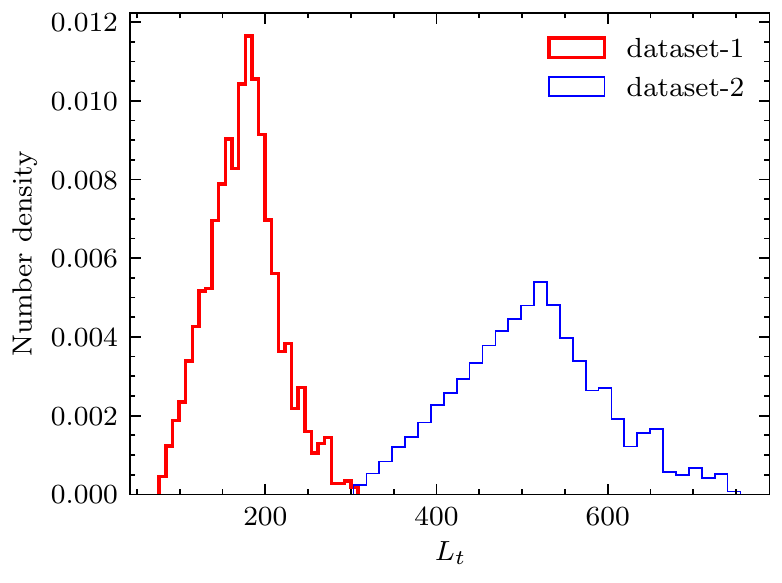}
\caption{target waveform length ($L_{t}$) distribution of the training sub-dataset in dataset-1 (thick red) and dataset-2 (thin blue). Note that the non-uniform distributions are caused by the parameter sampling and input-target split method described in Sec.~\ref{sec:data_prep}.}
\label{Fig:dstatistic}
\end{figure}

For divided target waveforms, we illustrate the number density distributions of waveform lengths in Fig.~\ref{Fig:dstatistic} (denoted by $L_{t}$). As shown in the figure, the distributions are not uniform. We reckon that this non-uniformity causes $L_{t}$-dependent accuracy of the DL model, which will be discussed in Sec.~\ref{sec:waveform validation}.

%The dashed green and solid blue lines respectively show. %input (inspiral) and target (merger and ringdown) waveforms. %Presented masses as annotation are chirp masses of the corresponding waveforms.

\section{Methods}
\label{sec:methods}

Since the duration of the GW emission within the detector's sensitive frequency band varies depending on the component masses or chirp mass of the binary system, we need a DL model capable of handling different size data. For this, we design a DL model with a novel architecture based on seq2seq, which is built for NLP. In this section, we briefly overview the original seq2seq model\footnote{For more details of the original model, we refer to ~\cite{cho2014learning,Sutskever:2014:SSL:2969033.2969173}.} and elaborate on our model below.

\begin{figure*}[t]
\centering
\includegraphics[width=.85\textwidth]{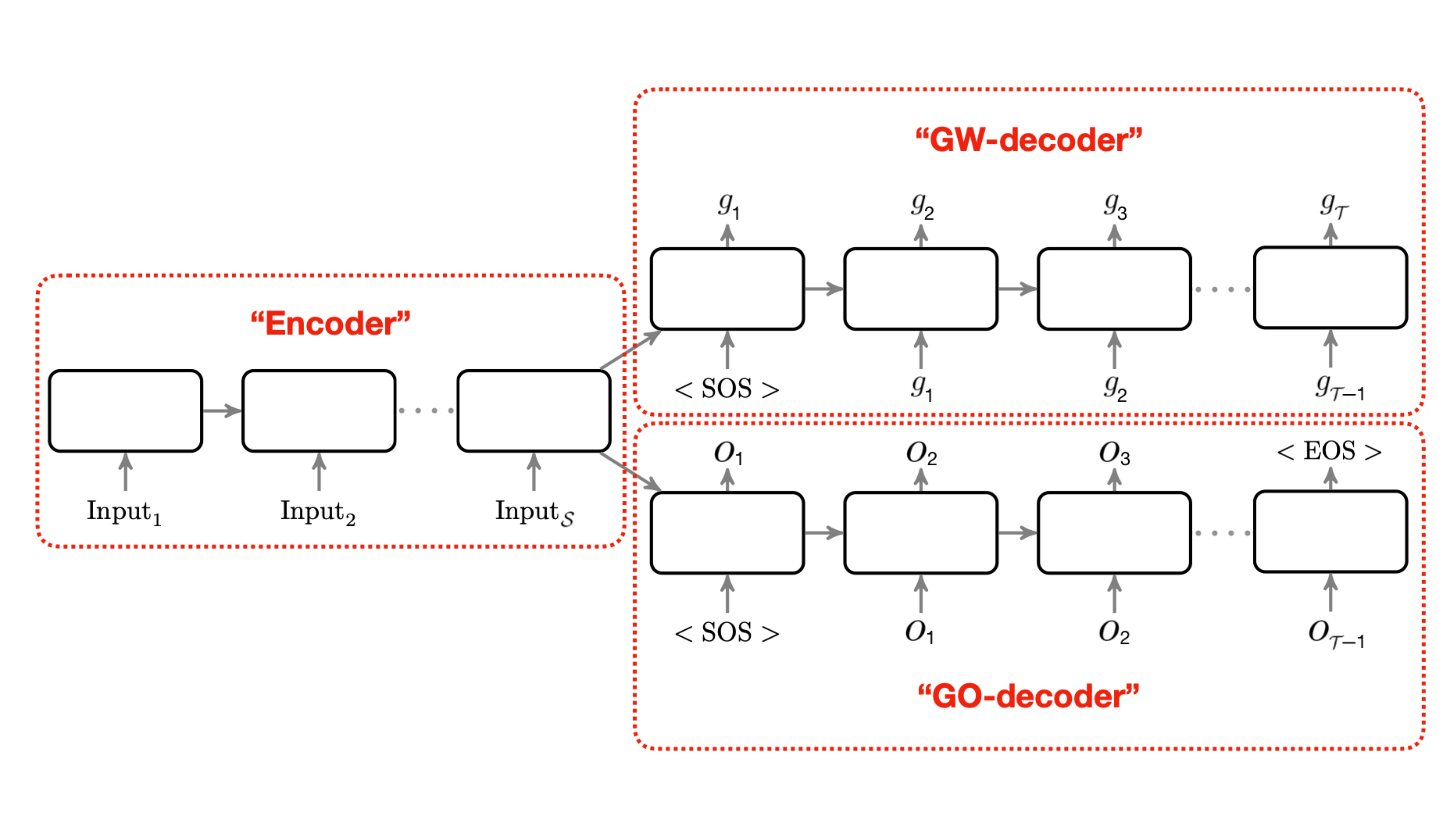}
\caption{\label{Fig:model_structure}The schematic workflow of the DDS2S model. The solid black boxes indicate RNN cells. The model sequentially takes $\mathcal{S}$ vectors as input waveforms and attempts to regenerate target waveforms and GO-function, $\cal{G}$. The decoders start computation when inputted $\langle$SOS$\rangle$ and retrieve $\mathcal{T}$ vectors as output waveforms until the GO-decoder yields a value under 0.5, marked by $\langle$EOS$\rangle$. Note that the decoders use the output of the previous computing-step as the input at the next computing-step. The detailed structural information of the model is tabulated in Table.~\ref{Tab:model}.}
\end{figure*}

\subsection{Original Sequence-to-Sequence Model}

DL models for NLP take a batch of sentences as inputs and output transformed sentences. For that, each word in the sentences should be digitized since machine learning models work numerically. With the linguistic property that the number of vocabularies in a specific language is limited to a finite number, each distinct word can be represented as a vector by word embedding~\cite{NIPS2000_1839}. Thus, the sentence prediction problem can be regarded as selecting words from a given dictionary. The vectorized sentences, however, have different sizes because every sentence is composed of a different number of words. 

To resolve the issue, the encoder, mapping the variable size input sequence into a fixed-size vector, is employed in the seq2seq model. Afterward, the transformed vectors, so-called representations, by the encoder are transmitted to the decoder, and it sequentially recovers the variable size target sentences. In the decoder calculation process, the output at the previous computing-step is taken as the input of the next step. Each sentence is required to end with the end-of-sequence token ($\langle$EOS$\rangle$), and the decoder starts and finishes its computation by taking and outputting $\langle$EOS$\rangle$. The conditional vector $\langle$EOS$\rangle$ can be defined differently depending on the user's preference.

\subsection{Dual-Decoder Sequence-to-Sequence Model}
\label{subsec:model}

In the work of the original model, Sutskever et al.~\cite{Sutskever:2014:SSL:2969033.2969173} were able to construct the $\langle$EOS$\rangle$, the interrupting condition of the decoder computation, using the linguistic property that the number of vocabularies is limited. Since the words in the dictionary can be discretely distinguished, it is clear to set the condition.

Regarding the GW-data, however, it becomes hazy to establish a criterion for interrupting the computing-step because the strain amplitudes of GWs are continuous real numbers: the number of possible cases is infinite, unlike the words in a dictionary. Thus, we cannot expect the model to produce an output that exactly matches a specific number by all digits. For example, when we set $\langle$EOS$\rangle=0$, the model is unlikely to obtain the exactly matching value in machine precision.

As a strategy for learning this continuous sequence, we design a modified seq2seq model (DDS2S, Fig.~\ref{Fig:model_structure}) with one encoder and dual-decoder, GW- and GO-decoder: the encoder encrypts input waveforms, GW-decoder recovers target waveforms, and GO-decoder predicts the length of the target waveforms. While the computational mechanisms of the encoder and decoders are identical to the ones in the original model, the approach for handling input and target data is different.

First, the input and target waveforms are divided into the number of $\mathcal{S}$ and $\mathcal{T}$ vectors with $\mathcal{R}$ elements. When ${\cal{R}} > 1$, the ends of the waveform elements are zero-padded before division to match the component numbers with the multiples of ${\cal{R}}$. The zero-padded lengths of input and target waveforms can be computed via $L_{s} = \mathcal{S} \mathcal{R}$ and $L_{t} = \mathcal{T} \mathcal{R}$~\footnote{Note that $L_{s}$ and $L_{t}$ are the lengths of input and target waveforms without zero-padding as $\mathcal{R} = 1$.}. Then, the encoder sequentially takes ${\cal{R}}$ elements of input waveforms ${\cal{S}}$ times and encrypts them into fixed-size vectors. The encoder outputs are transmitted to GW- and GO-decoders.

Subsequently, the GW-decoder regenerates $\mathcal{R}$ elements of target waveforms at every computing-step throughout the $\mathcal{T}$ step\footnote{The total computing-step multiplied by $\mathcal{R}$ and waveform length are compatible concepts, and one can convert them into the duration of GW by multiplying the inverse of the sampling rate, 4096Hz.}. The generated waveforms are stacked in the order of computing-step and compared with the target waveforms to calculate the error function. As the error function of the GW-decoder, ${\cal{I}}$, we use the sum of mean-squared error and negative cosine similarity;
\begin{equation}
    {\cal{I}}(g, t) = \frac{1}{{\cal{T}}} \Sigma_{i} (g_{i} - t_{i})^{2} - \frac{g \cdot t}{||g|| \, ||t||},
\end{equation}
where $g$ and $t$ are respectively the generated and target waveforms; ${\cal{T}}$ is the number of vectors for the given target waveform; $|| \cdot ||$ is $L^{2}$ norm.

Lastly, we establish the GO-function to endow the GO-decoder the capability to estimate the length of the target waveform. When the given target waveform consists of ${\cal{T}}$ vectors, we can set the integer condition, ${\cal{C}}$, for progressing from computing-step $\tau$ to $\tau+1$ as follows: 1 for proceeding and 0 for breaking.

\begin{equation}
    {\cal{C}}_\tau = 
    \begin{cases}
    1, & \mathrm{if} \quad 1 \leq \tau < {\cal{T}}-1 \\
    0, & \mathrm{if} \quad \tau \geq {\cal{T}}.
    \end{cases}
\end{equation}

\begin{figure}[t!]
\centering
\includegraphics[width=1.\linewidth]{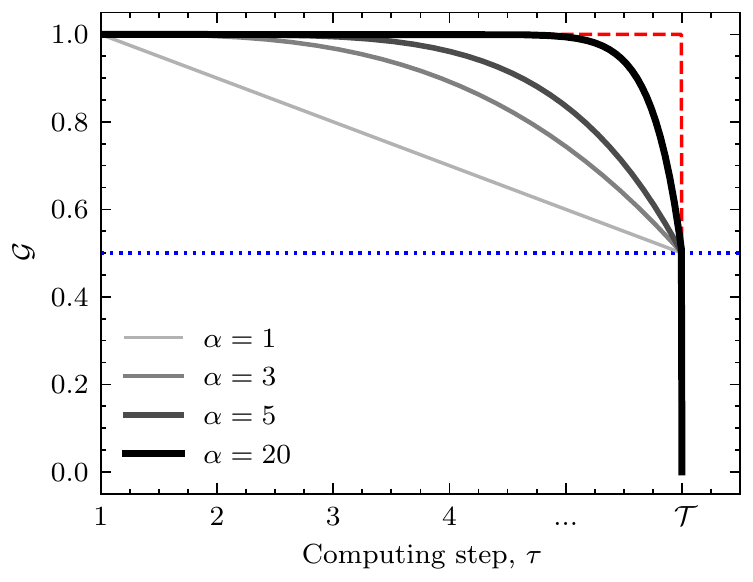}
\caption{The GO-function, $\cal{G}$, with several values of $\alpha$ in greyscale. The red dashed-line depicts how the integer condition, $\cal{C}$, changes according to the computing-step. As the value of the $\alpha$ increases, the function approximates the $\cal{C}$ values more accurately. We also draw the horizontal blue dotted-line at 0.5, the condition of interrupting decoders' computation.}
\label{Fig:go_func}
\end{figure}

We may use the set of $\cal{C_{\tau}}$ to train GO-decoder, but the discrete values and rapid decrease of ${\cal{C}}$ from $\tau={\cal{T}}-1$ to $\tau={\cal{T}}$ are inappropriate for the training of the DL model. Thereby, we define GO-function, ${\cal{G}}$, approximating the integer ${\cal{C}}$ values with a smooth decreasing pattern near $\tau={\cal{T}}$ and use the function to compute the mean-squared error with the GO-decoder outputs. The GO-function and the error function, ${\cal{J}}$, of the GO-decoder are described below.
\begin{equation}
    {\cal{G}}_\tau = 
    \begin{cases}
    1 - 0.5\left(\tau / {\cal{T}}\right)^{\alpha}, \quad & \mathrm{if} \quad 1 \leq \tau \leq {\cal{T}}-1 \\
    0, \quad & \mathrm{if} \quad \tau \geq {\cal{T}},
    \end{cases}
\end{equation}

\begin{equation}
    {\cal{J}}(o, {\cal{G}}) = \frac{1}{{\cal{T}}} \Sigma_{i} (o_{i} - {\cal{G}}_{i})^{2},
\end{equation}
where $o_{i}$ is the output of GO-decoder. Fig.~\ref{Fig:go_func} presents how the curve of the ${\cal{G}}$ varies according to different $\alpha$s. As the $\alpha$ is getting bigger, the GO-function approximates the ${\cal{C}}$ values more accurately. On the contrary, we find that the rapid decrease of ${\cal{G}}$ near $\tau={\cal{T}}$ hinders the training of the DL model when the $\alpha$ is too high. We empirically determine $\alpha$ of 5 for the training of the model.

The final loss for the training is the sum of the error function of GW- and GO-decoders, namely ${\cal{I}} + {\cal{J}}$. The model is trained by adjusting its parameters in such a way the error is minimized.

\begin{table*}[t!]
\centering
\caption{\label{Tab:model}Detailed structure of the DDS2S model.}
\begin{tabularx}{1.\linewidth}{l @{\extracolsep{\fill}} c c c}
\toprule
& Encoder & GW-Decoder & GO-Decoder \\ 
\colrule
{RNN cells} & GRU & GRU & GRU \\
{The number of RNN cells} & ${\cal{S}}$ & ${\cal{T}}$ & ${\cal{T}}$ \\
{The number of input layers} & 1 & 1 & 1 \\
{The number of hidden layers} & 4 & 4 & 4 \\
{The number of output layers} & - & 1 & 1 \\
{The number of input neurons} & ${\cal{R}}$ & ${\cal{R}}$ & 1 \\
{The number of hidden neurons} & 256 & 256 & 256 \\
{The number of output neurons} & - & ${\cal{R}}$ & 1 \\
{Activation function of input layers} & \texttt{Tanh} & \texttt{Tanh} & \texttt{Tanh} \\
{Activation function of hidden layers} & \texttt{Tanh} & \texttt{Tanh} & \texttt{Tanh} \\
{Activation function of output layers} & - & - & \texttt{Sigmoid}\\
\botrule
\end{tabularx}
\end{table*}

\begin{figure*}[tb!]
\includegraphics[width=.99\linewidth]{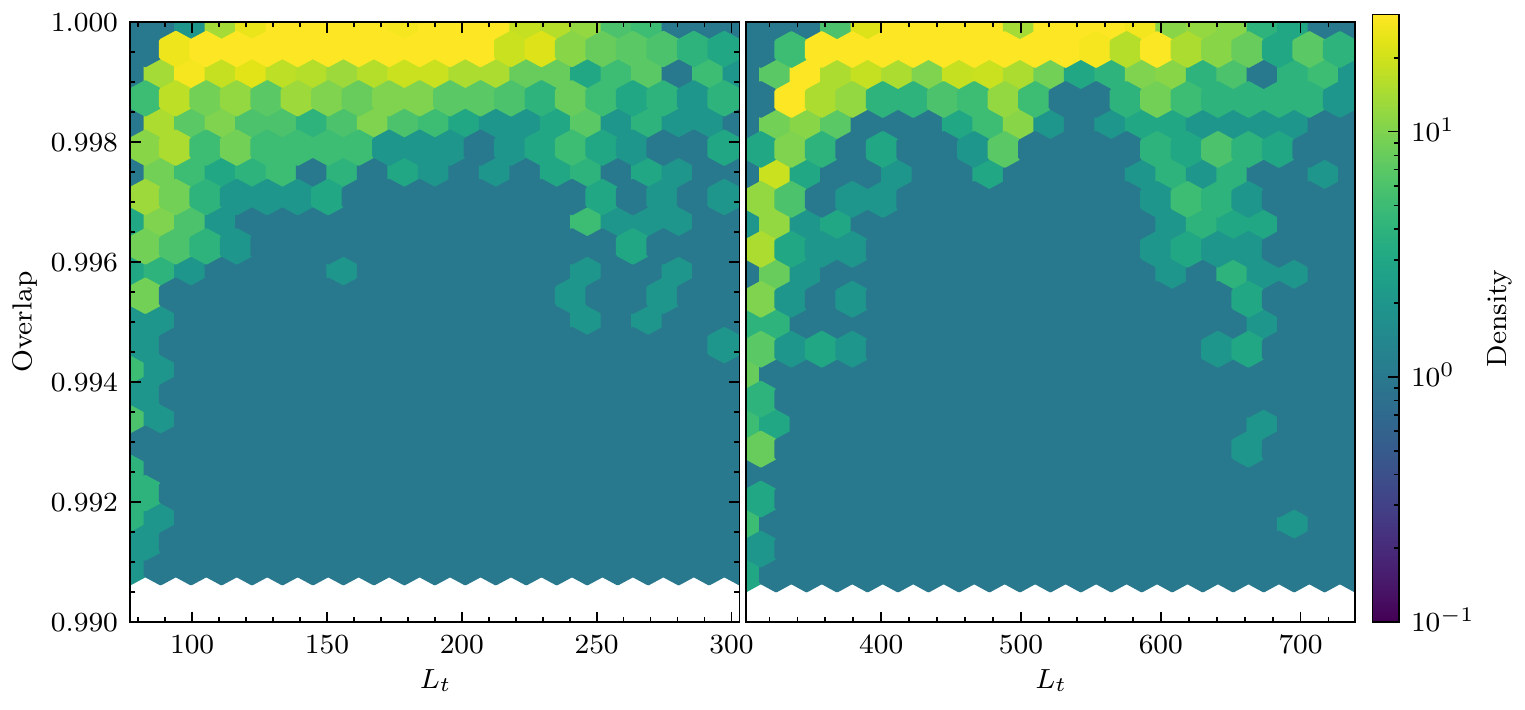}\label{Fig:overlap_scatter_1}
\caption{Density heatmap of overlap according to target waveform lengths, $L_{t}$, for the dataset-1 (left) and 2 (right). We draw the vertical axes of the two plots in the same range and scale. For clear contrast, we leave the regions with no samples empty at the bottom of the plots. As shown in the plots, overlaps of all the DL-waveforms are higher than 0.990. Besides, the averages of the waveforms from both datasets are over 0.999. However, a few shortest and longest samples have smaller overlap values. Considering the relatively small number of the shortest and longest waveforms in the training sub-dataset (Fig.~\ref{Fig:dstatistic}), it implies that the non-uniformity of the sub-dataset is related to the locally different accuracy of the DL model.}
\label{Fig:overlap_num_scatter}
\end{figure*}
\begin{figure*}[t!]
\subfloat[Best case of dataset-1, ${\cal{M}}$ = 0.999]{\hspace*{-1.cm}\includegraphics[width=.45\linewidth]{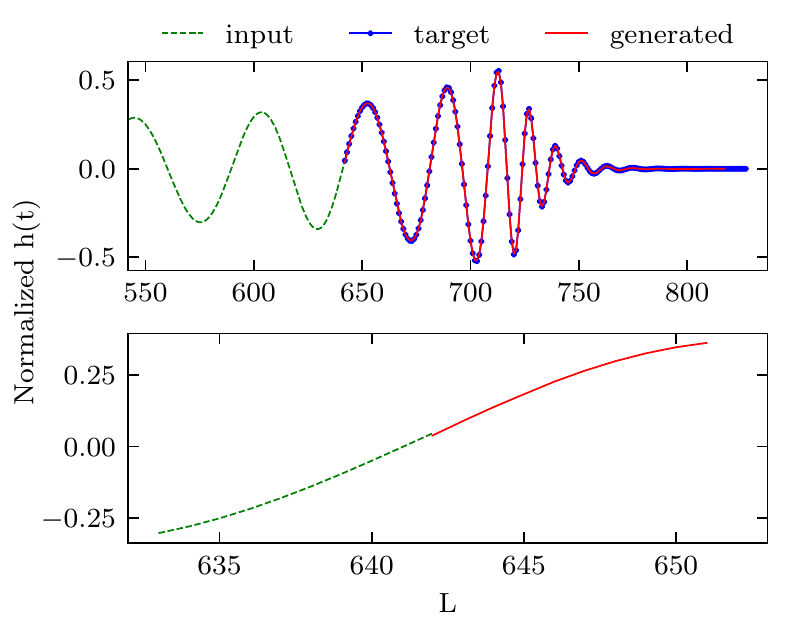}}
\hspace*{1cm}\subfloat[Best case of dataset-2, ${\cal{M}}$ = 0.999]{\hspace*{-1cm}\includegraphics[width=.45\linewidth]{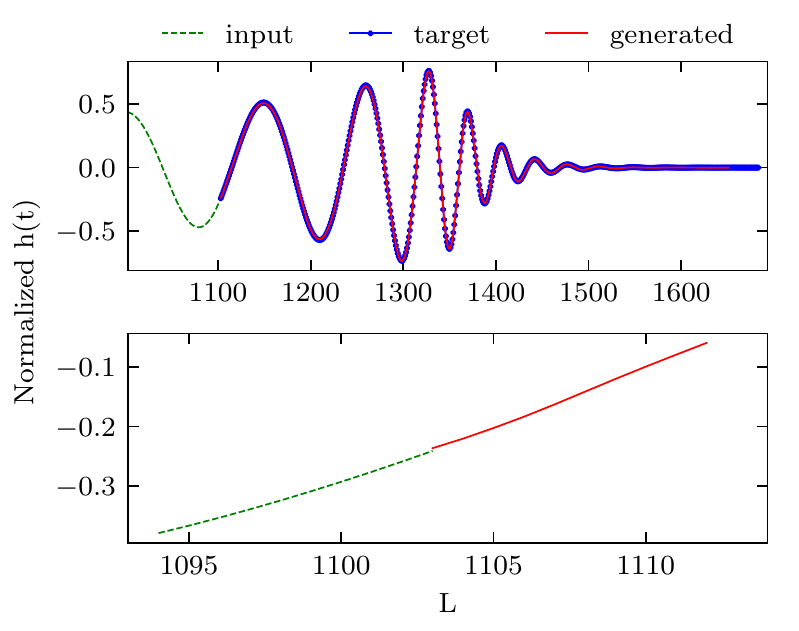}} \\
\subfloat[Worst case of dataset-1, ${\cal{M}}$ = 0.991]{\hspace*{-1cm}\includegraphics[width=.45\linewidth]{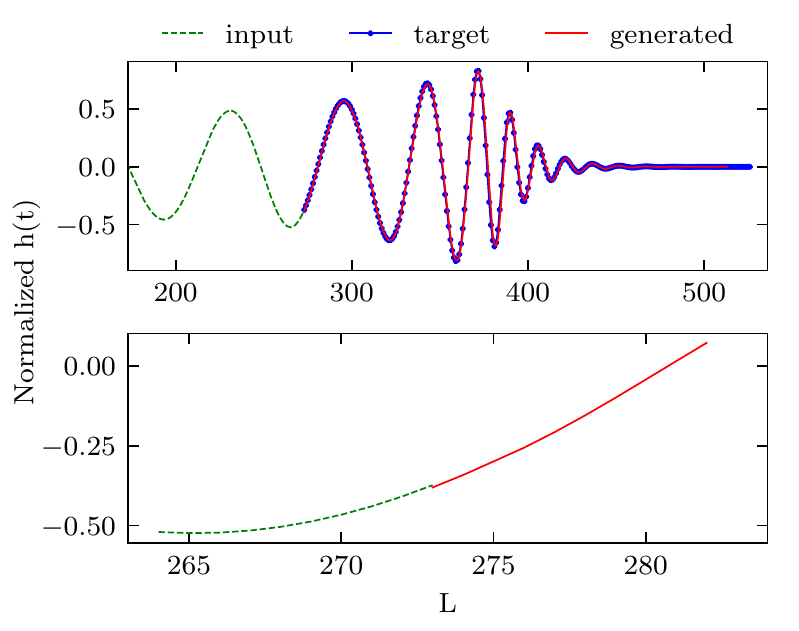}}
\hspace*{1cm}\subfloat[Worst case of dataset-2, ${\cal{M}}$ = 0.991]{\hspace*{-1cm}\includegraphics[width=.45\linewidth]{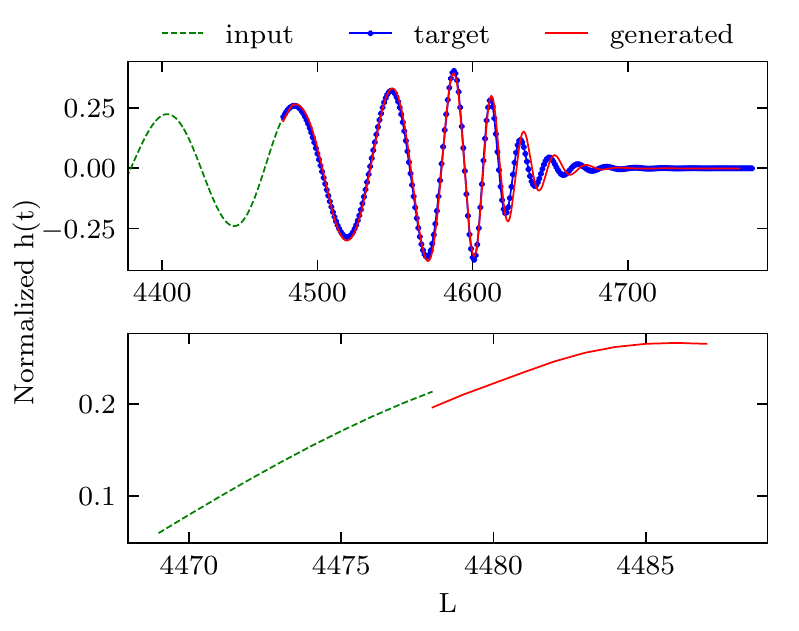}} \\
\caption{The input (green dashed), target (blue solid line with dots), and DL- (red solid) waveforms from dataset-1 (left column) and dataset-2 (right column) with the amplified images of connection points. The horizontal and vertical axes indicate the length of the waveforms in sampling unit and the normalized strain amplitude of the GWs, respectively. We only show a hundred sampling units of input waveforms in the plots for clear visualization. The top and bottom panels are the waveforms with the highest and lowest overlap cases, respectively.}
\label{Fig:generated_waveforms}
\end{figure*}

We apply the \texttt{Sigmoid} to the output layer of the GO-decoder since the GO-function should output values from 0 to 1. Then, we have given output values rounded to either 0 or 1. The computation continues when the rounded value is 1 and stops otherwise. Hence, the GO-decoder output below 0.5 serves as $\langle$EOS$\rangle$ in our case. For this reason, we define ${\cal{G}}$ to have a slightly higher value than 0.5 at $\tau={\cal{T}}-1$ because we expect the model to stop calculating at $\tau={\cal{T}}$. For the DDS2S model, we newly define zero vectors with $\mathcal{R}$ elements as a start-of-sequence token ($\langle$SOS$\rangle$), which is inputted at the start of decoder computation.

Among the prominent RNN cells, we choose Gated Recurrent Unit~(GRU)~\cite{cho2014learning} for the encoder and both decoders because the setting with GRU showed higher accuracy and faster training than Long-Short Term Memory~\cite{hochreiter1997long,gers1999learning}, another well-known RNN cell. A fully connected layer is placed at the end of the decoders' hidden layers to convert hidden states to vectorized outputs with ${\cal{R}}$ components. We use the hyperbolic tangent as the activation function for hidden layers of each RNN cell of encoder and decoders.

For the model structure, we find an empirically optimal model configuration varying the number of neurons in hidden layers (hereafter, hidden neurons) based on the overlap to a reference waveform, which we will discuss in the following sub-section. The information on the network configurations and hyperparameters of the optimal model is summarized in Table~\ref{Tab:model}.

\subsection{Overlap}
\label{subsec:overlap}
We use overlap to assess the DL-waveforms' accuracy. The normalized overlap, $\cal{M}$, of the DL-waveform $g$ and the target $t$ can be computed as
\begin{align}
\cal{M} &\equiv \frac{(g|t)}{\sqrt{(g|g)(t|t)}}, \label{eq:overlap}
\end{align}
where $(g|t) = \int_{-\infty}^{\infty} \tilde{g}(f)\tilde{t}^{*}(f) df$. $\tilde{g}$ and $\tilde{t}$ are the Fourier transform of $g$ and $t$, respectively, and asterisk mark (*) is complex conjugate. Note that $\cal{M}$ becomes 1 for the perfect match and 0 for the perfect mismatch between $g$ and $t$. 

From the grid-search described in Appendix~\ref{Append:AccNeurons}, we choose an empirically optimal model configuration, maximizing the minimum overlap of the model's output waveforms. Providing accuracy, we use the setup with 256 hidden neurons and ${\cal R} = 1$. Henceforward, we shall only discuss the results of the model with 256 hidden neurons and ${\cal R} = 1$. The detailed explanation can be found in Appendices~\ref{Append:AccNeurons} and~\ref{Append:TimeAccR}.

\section{Results}
\label{sec:results}

\subsection{Waveform Validation}
\label{sec:waveform validation}

The Fig.~\ref{Fig:overlap_num_scatter} depicts the overlap density heatmap between the DL-waveforms and corresponding target EOB-waveforms of the dataset-1 and 2. All of the DL-waveforms are in excellent agreement with their target waveforms in both cases as the minimum value of overlaps is higher than $0.990$\footnote{For comparison, the authors of Ref.~\cite{Sturani_2010} have shown that the overlap between numerical and their phenomenological waveforms ranges from 0.95 to 0.99. On the other hand, Ref.~\cite{2020PhLB..80035081W} have shown their model results in the overlap $\geq 0.99$.}. Furthermore, the mean overlaps of waveforms from both datasets are higher than $0.999$, indicating less than 0.1\% average error.

However, as we can see from the figure there are several outliers whose overlaps are substantially smaller than the majority. We explore the dependence of the overlap on the target waveform length to track down possible reasons for relatively poor overlap cases. The heatmap shows the distribution of the overlaps concerning the length of the target waveforms. The overlap of dataset-1 (dataset-2) tends to decrease at the short-end and long-end of the target waveform length, i.e., $L_{\mbox{t}} \lesssim 100$ or $L_{\mbox{t}} \gtrsim 250$ ($L_{\mbox{t}} \lesssim 400$ or $L_{\mbox{t}} \gtrsim 600$). As shown in Fig.~\ref{Fig:dstatistic}, the training samples in the range of $100 \lesssim L_{\mbox{t}} \lesssim 250$ of dataset-1 and $400 \lesssim L_{\mbox{t}} \lesssim 600$ of dataset-2 dominate the number distribution of the target waveform length. It can be attributed to the fact that the model is more likely to weigh the majority of the training sub-dataset.

We also visually inspect the agreement between the DL-waveforms and target waveforms. Fig.~\ref{Fig:generated_waveforms} shows the best and worst overlap cases of the DL-waveforms. The overlaps of the best cases for both datasets are ${\cal M}=0.999$. The time-series of the DL-waveforms matches well with the target waveforms. For the worst cases, the overlaps of the two datasets are both 0.991 (Fig.~\ref{Fig:generated_waveforms}~(c) and (d)). We see that there exist small discontinuities between the DL- and input waveforms as shown in the lower panel of the figure. We may resolve the discontinuity by post-processing or letting the DL model generate the whole waveform in the inspiral-merger-ringdown phase at once. We leave this issue to future work.

\subsection{Injection Test}
\label{sec:injection tests}

Next, we attempt to use the DL-waveform templates in simplistic search of parameters, i.e., $m_1$, $m_2$, and the event time of the simulated GW signals. To replicate practically used waveform templates, we hybridize inspiral \texttt{SEOBNRv4}-waveform and merger-ringdown DL-waveform by simply concatenating the two waveforms. One may implement sophisticate hybridization of waveforms, but it is beyond the scope of this work. We perform parameter grid-search instead of Markov Chain Monte Carlo, typically executed for the parameter estimation of GWs~\cite{van2008parameter}, due to the practical difficulty of plugging a new waveform model in the existing parameter estimation code~\cite{PhysRevD.88.062001}. For the computation of SNR and the search of the events, the matched filtering engine of \texttt{PyCBC}~\cite{Usman:2015kfa} is used.

To simulate the observation data embracing a GW signal, we use the LIGO-Hanford O1 data provided by GW Open Science Center\footnote{https://www.gw-openscience.org/archive/O1/}. We randomly select a 32-second segment from the data without any known GW signals and inject a \texttt{SEOBNRv4}-waveform into the center. While we use five sets of different injection parameters and distances, we fix the inclination angle to 30\degree for simplicity. The configuration setups of the tests are tabulated in the first three columns of Table~\ref{Tab:Mass_estimation}.

By performing the parameter grid-search for multiple injection waveforms, we retrieve injection parameters in all examinations within the 90\% confidence interval. We first define the search parameter sets, ($m_{1}$, $m_{2}$) on regularly-spaced grid of the parameter space. Then, we construct the full IMR waveform templates by hybridizing the inspiral waveform and the merger-ringdown DL-waveform using \texttt{SEOBNRv4} and DDS2S, respectively, for the parameter sets. Across the parameter sets, we compute SNR by matched filtering with each waveform template using \texttt{PyCBC} on the simulated data. Assuming the likelihoods of the parameter sets are proportional to the SNR, we estimate the probability density function (PDF) of the parameters. Then, we marginalize the PDF with respect to each parameter and acquire the median as the best-fit parameters with their 90\% confidence interval. Subsequently, we repeat the entire process with different combinations of injection masses and distances. The best-fit parameters with confidence intervals and their SNRs are summarized in the last two columns of Table~\ref{Tab:Mass_estimation}.

\begin{figure}[t]
\subfloat{\includegraphics[width=.95\linewidth]{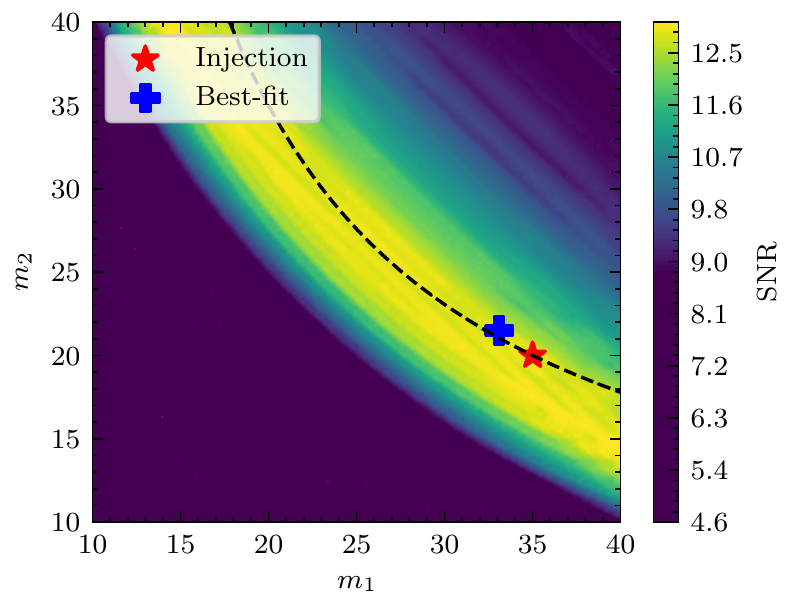}}
\caption{\label{Fig:fcontour}Filled contour map of SNR in the parameter space for the injection signal with $m_1=35M_{\odot}$ and $m_2=20M_{\odot}$. Each of the red star and blue plus markers indicates injection and best-fit parameters. The black dashed line is a contour with the level of injection chirp mass. The best-fit parameters and the high SNR region arise in the vicinity of the contour line. Although our parameter space is restricted with the condition $m_{1} \geq m_{2}$, the filled contour map is reflected on the slope of 1 line for aesthetic visualization.}
\end{figure}

The best-fit parameters and the high SNR region emerge around the chirp mass contour line of the injected signal. Since the chirp mass of GW is governed by the frequency and frequency derivative~\cite{Abbott:2016bqf}, and its SNR depends on frequency evolution~\cite{1998PhRvD..57.4535F}, the SNR of GW again relies on the chirp mass. It is well-reflected in the example contour map of the signal with $m_1=35M_{\odot}$ and $m_2=20M_{\odot}$ (Fig.~\ref{Fig:fcontour}).

\begin{figure}[t]
\subfloat{\includegraphics[width=1.\linewidth]{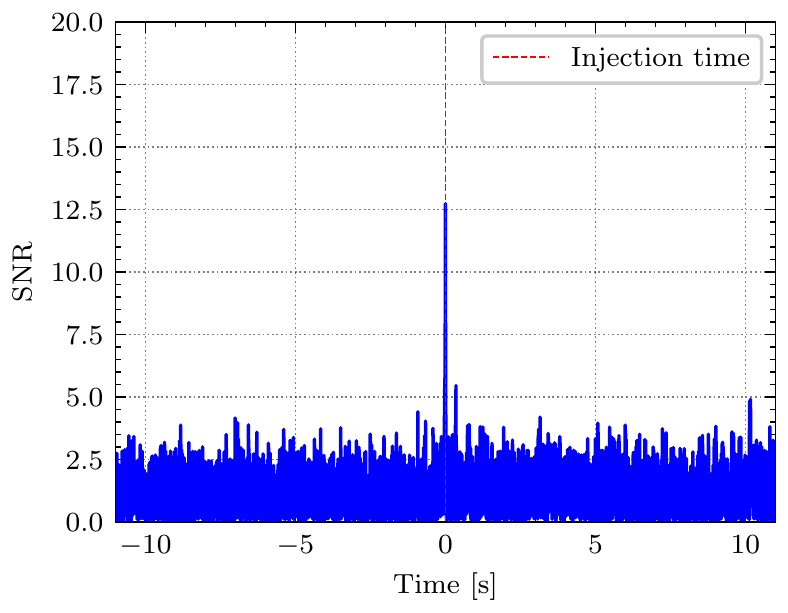}}
\caption{\label{Fig:search_snr}SNR time-series computed by matched filtering engine of \texttt{PyCBC} and best-fit DL-waveform template of Fig.~\ref{Fig:fcontour}. The injected signal is the \texttt{SEOBNRv4}-waveform ($m_1=35M_{\odot}$ and $m_2=20M_{\odot}$). Here, we initialize the start time of the injected signal to 0, marked by the red dashed line. Note that the SNR peak occurs at the injection time.}
\end{figure}

Using the best-fit parameters found from the grid search, we perform event time searches and find the SNR peak at where we inject the signals. We illustrate SNR time-series of the above example case in Fig.~\ref{Fig:search_snr}. As can be seen in the figure, the peak SNR occurs at the center of the data segment, where we have injected the simulated signal.

% \textcolor{brown}{Furthermore, we naively estimate the impact of the systematic error of the DL-based waveforms in the parameter estimation by performing the grid-search for the set of injection parameters shown in Fig~\ref{Fig:fcontour}. The systematic error is obtained by the offset of the best-fit parameters from the true injected ones. We compare it with the statistical errors of the same parameter as increasing the SNR of the injected signal and find that the magnitude of the systematic error becomes comparable to the 1-$\sigma$ statistical error at $\mathrm{SNR} \sim \mathcal{O}(10)$. For reference, it is rigorously discussed that the magnitude of the systematic errors from 3.5PN (post-Newtonian approximation of order 3.5) waveform with ${\cal M} > 0.9999$ commensurate with the SNR $\sim 1000$ statistical errors~\cite{PhysRevD.76.104018}.}

It is known that the systematic error from waveform approximants is independent of SNR, while the statistical error due to noise roughly scales as $1/\mathrm{SNR}$. One can readily expect that the systematic error could dominate in higher SNR signals. Cutler and Vallisneri~\cite{PhysRevD.76.104018} have presented  rigorous computation of the systematic errors in parameter estimation using 3.5PN (post-Newtonian approximation of order 3.5) waveforms for inspiral signals of massive black hole binaries. They have shown that the magnitude of the systematic errors from 3.5PN waveforms with ${\cal M} > 0.9999$  commensurate with the SNR $\sim 1000$ statistical errors. Motivated by this, we roughly estimate the impact of systematic error of our DL-based waveform on the parameter estimation by repeating the grid-search of parameters as described above with varying SNR of the injected signal. By comparing the systematic error with the statistical errors of the same parameter as increasing the SNR of the injected signal, we find that the magnitude of the systematic error becomes comparable to the 1-$\sigma$ statistical error at $\mathrm{SNR} \sim \mathcal{O}(10)$ in our DL-based waveform approximant.\footnote{Note that our approach for finding the SNR level where the two errors become similar is not rigorous. For a more in-depth exploration of the systematic errors, refer to~\cite{PhysRevD.76.104018}.}

\subsection{Performance Dependence on the Dataset Size}
\label{sec:additional experiments}

We inspect the dependence between the accuracy of the DL model and the number of waveforms in the training sub-dataset. The test is performed to explore the viability of applying the proposed model to NR-waveforms, in which only a few thousands exist~\cite{Boyle:2019kee,healy2019second,Jani_2016,Healy_2017}. We generate four reduced datasets with half and the tenth number of waveforms in the original training data of dataset-1 and -2, maintaining the number of waveforms in the validation and test data.

\begin{table*}[t!]
\centering
\caption{\label{Tab:Mass_estimation}Summarized results of the injection tests. The best-fit parameters and their SNR for the injected signals are computed by \texttt{PyCBC} matched filtering engine with DL waveform templates. We establish template waveforms by hybridizing inspiral \texttt{SEOBNRv4} and merger-ringdown DL waveforms. The $m_1$ and $m_2$ are given in the unit of the solar mass. I, M, and R indicate inspiral, merger, and ringdown phases, respectively.}
\begin{tabularx}{\linewidth}{c @{\extracolsep{\fill}} c c c c c c}
\toprule

Template approximant & Distance (Gpc) & Injection ($m_1$, $m_2$) & Best-fit ($m_1$, $m_2$)  & SNR \\
\colrule
\multirow{5}{*}{EOB (I) + DL (MR)} & $1.6$ & $80.0$, $65.0$ & $80.1_{-14.5}^{+13.7}$, $61.7_{-16.4}^{+18.3}$ & $14.5$  \\
                                   & $1.5$ & $70.0$, $60.0$ & $73.9_{-16.9}^{+16.5}$, $58.6_{-14.4}^{+16.6}$ & $13.0$ \\
                                   & $0.8$ & $35.0$, $20.0$ & $33.1_{-6.8}^{+5.6}$, $21.5_{-8.4}^{+9.0}$ & $12.7$ \\
                                   & $0.7$ & $30.0$, $25.0$ & $31.6_{-7.0}^{+6.3}$, $22.7_{-8.3}^{+8.6}$ & $15.3$ \\
                                   & $0.6$ & $25.0$, $20.0$ & $28.3_{-8.4}^{+8.0}$, $18.9_{-6.6}^{+7.1}$ & $15.7$ \\
\botrule
\end{tabularx}
\end{table*}

We find that one-tenth of the original size is enough to reach the required accuracy of $\mathcal{M} \geq 0.99$. The model is trained more than five times with each reduced training data. It turns out that the minimum and average values of overlap are higher than 0.990 and 0.999, equivalent to 1.0\% and 0.1\% error, respectively, for all DL-waveforms of the trained model from each run. The mean values for the averaged overlaps and minimum overlaps from more than five individual runs are tabulated in Table~\ref{Tab:size_test}. We also present the results of Sec.~\ref{sec:waveform validation} for comparison in the last column. The relative dataset size in the table means the ratio of the number of waveforms in the training data to the number of waveforms in the original training data. The result shows that reducing the number of waveforms down to 1000 for the training hardly affects obtaining the desired accuracy. Hence, we advocate that the application of the DDS2S model to NR-waveforms is feasible.

\begin{table}[t!]
\centering
\caption{\label{Tab:size_test}Accuracy variation of the DL model according to dataset size. We also show the results of Sec.~\ref{sec:waveform validation} in the last column of the table for comparison. The mean values for the minimum and average overlaps from more than five individual runs for each dataset are summarized in the table. The value of the relative dataset size is the ratio of the number of waveforms between the reduced training sub-dataset and the sub-dataset introduced in Sec.~\ref{sec:data_prep}.}
\begin{tabularx}{\linewidth}{c @{\extracolsep{\fill}}c c c}
\toprule
\multicolumn{1}{c}{Relative dataset size} & 0.1 & 0.5 & 1\\ 
\colrule
\multicolumn{1}{c}{Minimum overlap (dataset-1)} & \multicolumn{1}{c}{0.991} & \multicolumn{1}{c}{0.990} & \multicolumn{1}{c}{0.991}\\
\multicolumn{1}{c}{Minimum overlap (dataset-2)} & \multicolumn{1}{c}{0.990} & \multicolumn{1}{c}{0.990} & \multicolumn{1}{c}{0.991}\\
\multicolumn{1}{c}{Average overlap (dataset-1)} & \multicolumn{1}{c}{0.999} & \multicolumn{1}{c}{0.999} & \multicolumn{1}{c}{0.999}\\
\multicolumn{1}{c}{Average overlap (dataset-2)} & \multicolumn{1}{c}{0.999} & \multicolumn{1}{c}{0.999} & \multicolumn{1}{c}{0.999}\\
\botrule
\end{tabularx}
\end{table}

%  in the training sub-dataset to that of waveforms in the original training sub-dataset.

\section{Summary and Discussion}
\label{sec:discussion}

The efficiency of matched filtering for searching GW signals buried in noisy GW data has been proved by recent successful detections of GW signals. Although NR can increase the accuracy of template waveforms, expensive computational costs of running NR limit the use of it for the generation of a sufficiently large number of template waveforms. This drawback of NR eventually led to the use of approximate waveforms for the matched filtering instead. Motivated by such difficulties, we have examined the DL method for the generation of template waveforms with much smaller computational costs but comparable accuracy to NR.

To study the feasibility of this consideration, we have implemented the DDS2S model. The encoder-decoder structure is capable of handling the variable sizes of different waveforms, and the dual-decoder structure enables the model to control the continuous real-numbered sequences.

We also have examined the applicability of the waveforms by computing the overlap with EOBNR-based waveforms and performing the injection test. The accuracy of the DL-based waveforms is found to be better than 99.9 \% in most combinations of the masses, while a small number of outliers with overlap as small as 0.99 exists. 
%We also have found that the systematic errors of our waveform templates stay smaller than statistical errors originated from instrumental and background errors within the range of SNR $\lesssim 400$. 
In the injection test, we have recovered the event time of waveforms injected into real noise data with the conventional matched filtering engine of \texttt{PyCBC}.

We have found that the method generating merger-ringdown waveforms using the inspiral waveforms needs to be improved. For example, we have seen that discontinuities occurred between input and output waveforms, as shown in Fig.~\ref{Fig:generated_waveforms}, although the minimum overlap of DL-waveforms to the EOB-waveform was higher than 0.990. To avoid this issue, we may take a new strategy of generating a full IMR waveform. However, the main goal of this paper is to demonstrate the feasibility of adopting DL to model the merger-ringdown waveforms. Hence, we leave the implementation of a DL model generating the full waveforms to future work.

Regarding the speed of waveform generation, the DDS2S model has an advantage over other waveform approximants when computing a batch of multiple waveforms simultaneously. For computing a single waveform, EOB is faster than the DDS2S model, typically taking $\mathcal{O}(10^{-2})$~seconds using a modern CPU core. However, the DDS2S model generates $\sim 1500$ waveforms using pre-generated inspiral waveforms in $\mathcal{O}(1)$~seconds using NVIDIA GeForce GTX 1080, while EOB took $\mathcal{O}(10)$~seconds. The disparity arises since the DL models are specialized for batch computations, which process multiple data at once.

The DDS2S model has been built to learn how to predict the output waveforms only from the given input waveforms without any specific physical information of the source binary system. Thus, we can readily extend this work to various systems of interest. 

For a more precise description of realistic physical binary systems, we need to have waveform models for more complex binaries: a wider range of the mass ratios, the spin of each component, eccentricity of the orbits. GWs from unbound orbit such as hyperbolic and parabolic encounters are also of great interest. Lastly, it is worthwhile to mention that recalibration of full IMR waveforms to increased amounts of NR waveform data is in progress in the community.~\cite{LVKWhitePaper}

Our approach described in this paper can potentially be applied to more complex systems described above because DDS2S only depends on training data, not any assumptions or approximations on which other waveform models are based. Moreover, we have observed that $\sim 1000$ training waveforms are sufficient for the model to reach the expected level of accuracy in Sec.~\ref{sec:additional experiments}. Thus, as long as there is a sufficiently large number of  training waveform samples for any systems or NR are given, DDS2S can be trained to generate accurate waveforms in principle.

\begin{acknowledgments}

The authors are grateful to Chunglee Kim and Hyung Won Lee for their helpful comments. And the authors also thank Young-Min Kim, Hee-Suk Cho, and Whansun Kim for fruitfully discussion. G.C is supported by the ERC Consolidator Grant ``Precision Gravity: From the LHC to LISA" provided by the European Research Council (ERC) under the European Union's H2020 research and innovation programme, grant agreement No. 817791. This work was supported by the National Research Foundation of Korea (NRF) grant funded by the Korea government (MSIT) (No. 2019R1A2C2006787). This work is partially supported by the Global Science experimental Data hub Center (GSDC) at KISTI. This research was also partially supported by National Institute for Mathematical Sciences (NIMS) funded by MSIT (B19720000) and the NRF grant funded by the MSIT (NRF-2020R1C1C1005863, NRF-2020R1I1A2054376). This research has made use of data, software and/or web tools obtained from the Gravitational Wave Open Science Center (https://www.gw-openscience.org), a service of LIGO Laboratory, the LIGO Scientific Collaboration and the Virgo Collaboration. LIGO is funded by the U.S. National Science Foundation. Virgo is funded by the French Centre National de Recherche Scientifique (CNRS), the Italian Istituto Nazionale della Fisica Nucleare (INFN) and the Dutch Nikhef, with contributions by Polish and Hungarian institutes. This paper has the LIGO document number LIGO-P1900207.
\end{acknowledgments}

\appendix 
\setcounter{figure}{0}
\setcounter{table}{0}
\renewcommand{\thefigure}{\Alph{section}.\arabic{figure}}
\renewcommand{\thetable}{\Alph{section}.\arabic{table}}

\section{Empirically Optimal Number of Hidden Neurons \label{Append:AccNeurons}}

We investigate the influence of the hidden neurons on the accuracy of the models; 64, 128, and 256 hidden neurons.

Accuracy-wisely, we find that the model with 256  hidden neurons is most suitable amid the tested cases. To compare model accuracy according to the number of hidden neurons, minimum and average overlap between DL-waveforms and corresponding target waveforms are computed. Table~\ref{Tab:model size test} summarizes the minimum and average overlaps of the models for dataset-1 and -2. The minimum overlap values of each model from dataset-1 (dataset-2) are 0.984, 0.990, and 0.991 (0.977, 0.989, and 0.991) in the increasing order of the model size. All of the average overlaps are the same as 0.999, except the case of the smallest model with dataset-2, whose overlap is 0.998 (overlaps of 0.999 and 0.998 are equivalent to $0.1\%$ and $0.2\%$ errors). Namely, the model with 256 hidden neurons shows the highest accuracy.

\begin{table}[h]
\centering
\caption{\label{Tab:model size test}Minimum and average overlap values of the test sub-dataset in dataset-1 and -2 according to models with the different number of hidden neurons.}
\begin{tabularx}{1.\linewidth}{l @{\extracolsep{\fill}} c c c}
\toprule
The number of hidden neurons & 64 & 128 & 256 \\ 
\colrule
Minimum overlap (dataset-1) & 0.984 & 0.990 & 0.991 \\
Minimum overlap (dataset-2) & 0.977 & 0.989 & 0.991 \\
Average overlap (dataset-1) & 0.999 & 0.999 & 0.999 \\
Average overlap (dataset-2) & 0.998 & 0.999 & 0.999 \\
\botrule
\end{tabularx}
\end{table}

\section{Computing Time and Accuracy Variation of The Model According To ${\cal{R}}$ \label{Append:TimeAccR}}
We examine how the number of elements ${\cal {R}}$ in an RNN cell affects the model in the aspects of computing time and accuracy. Table~\ref{Tab:grid_search_2} tabulates the typical elapsed time with a minimum overlap of each case on dataset-1 and -2. Although the model can speed up by increasing ${\cal{R}}$, the accuracy expense renders the model inapplicable for practical use.

\begin{table}[h]
\centering
\caption{Computation time and overlap variation with respect to the number of elements, $\mathcal{R}$, in a RNN cell. \label{Tab:grid_search_2}}
\begin{tabularx}{1.\linewidth}{c @{\extracolsep{\fill}} c c c c}
\toprule
\multirow{2}{*}{$\cal{R}$} & \multirow{2}{*}{$T_{1}$} & \multirow{2}{*}{$T_{1500}$} & \multicolumn{2}{c}{Minimum overlap} \\
& & & dataset-1 & dataset-2 \\ 
\colrule
1 & ${\cal{O}}(10^{-1})$ & ${\cal{O}}(1)$ & 0.991 & 0.991 \\
10 & ${\cal{O}}(10^{-2})$ & ${\cal{O}}(10^{-1})$ & 0.913 & 0.910 \\
100 & ${\cal{O}}(10^{-3})$ & ${\cal{O}}(10^{-2})$ & 0.823 & 0.805 \\
\botrule
\end{tabularx}
\end{table}

\bibliographystyle{apsrev4-1}
\bibliography{./References_GW.bib,./References_ML.bib}

\onecolumngrid
 
\newcommand\Tstrut{\rule{0pt}{3ex}}       % "top" strut
\newcommand\Bstrut{\rule[-1.5ex]{0pt}{0pt}} % "bottom" strut
\newcommand{\TBstrut}{\Tstrut\Bstrut} % top&bottom struts

\end{document}